\documentclass[fleqn,12pt,twoside]{article}
\usepackage{espcrc1}

\usepackage{graphicx}
\usepackage[figuresright]{rotating}

\newcommand{\AmS}{{\protect\the\textfont2
  A\kern-.1667em\lower.5ex\hbox{M}\kern-.125emS}}

\title{In-medium effects on the $K^-/K^+$ ratio at GSI}

\author{L. Tol\'os\address[MCSD]{Institut f\"ur Theoretische Physik,
J. W. Goethe-Universit\"at\\ D-60054 Frankfurt am Main,
Germany} \address{ Departament d'Estructura i Constituents de
la Mat\`eria,\\ Universitat de Barcelona, Diagonal 647, 08028
Barcelona, Spain}
\thanks{e-mail:tolos@th.physik.uni-frankfurt.de}
\thanks{AvH Fellow},
A. Polls\addressmark,
 A. Ramos\addressmark \, and
J. Schaffner-Bielich\addressmark[MCSD]}

\begin{document}

% typeset front matter
\maketitle

\begin{abstract}
The in-medium  modifications  on the  $K^-/K^+$ ratio produced  at GSI
are studied.  Particular  attention is paid to the 
properties  of antikaons, which  determine  the chemical potential and
       temperature at  freeze-out conditions.
 Different approaches  have  been considered:  non-interacting
$K^-$, on-shell self-energy and  single-particle spectral  density. We
observe  that the full off-shell approach  to the spectral
density reproduces the Brown et al. `broad-band equilibration' which is
crucial to explain an enhanced $K^-/K^+$ ratio.
\end{abstract}

\section{Introduction}

The medium modifications of mesons with strangeness such as kaons and antikaons can be studied in connection to heavy-ion experiments for energies around  1-2 AGeV \cite{Oeschler}. One surprising observation in C$+$C and Ni$+$Ni collisions
\cite{Barth} is that
 the
$K^-$ multiplicity and that of $K^+$ are of the same order of magnitude although in
 $pp$ collisions  the  $K^+$ multiplicity exceeds the $K^-$ one by 1-2 orders of magnitude at the same energy above threshold.
Another interesting observation is that the $K^-/K^+$ ratio stays almost
constant for C$+$C, Ni$+$Ni and Au$+$Au collisions for 1.5 AGeV
\cite{Barth}. 
Both observations could be interpreted to be a manifestation of an 
attractive $K^-$ optical
potential.
On the other hand, equal centrality  dependence  for the $K^+$  and  $K^-$
mesons has also been observed in Au$+$Au and  Pb$+$Pb reactions at 1.5
AGeV \cite{Barth}. Actually, the independence of centrality of the
$K^-/K^+$ ratio was claimed to indicate that no in-medium effects were needed in order to explain the experimental ratio \cite{Cleymans}.
However, the concept of ``broad-band equilibration'' was introduced by Brown
et al. \cite{Brown} in order to explain the centrality independence but including medium modifications of antikaons.

In this work we study the implications of introducing the $K^-$ spectral density for the $K^-/K^+$ ratio in order to address the above mentioned issues.

\section{In-medium modifications on the \mbox{\boldmath$K^-/K^+$} ratio}

We present a brief description of the statistical models which are applied for the calculation of the  $K^-/K^+$ ratio. 
Statistical models are based on
the assumption   that  the  particle ratios in    relativistic  heavy-ion
collisions can be described by  two parameters, the baryonic chemical potential $\mu_B$ and the temperature $T$ \cite{Cleymans}.

Therefore, by using  canonical strangeness conservation 
and taking into account the most relevant contributions
 in the $S=0,\pm 1$ sectors, the inverse ratio $K^+/K^-$ is given by
\cite{Cleymans,Laura03}
\begin{eqnarray}
\frac{K^+}{K^-}
=\frac{Z^1_{K^+}(Z^1_{K^-}+Z^1_{\Lambda}+
Z^1_{\Sigma}+Z^1_{\Sigma^*})}{Z^1_{K^-}Z^1_{K^+}}
=1+\frac{Z^1_{\Lambda}+Z^1_{\Sigma}+Z^1_{\Sigma^*}}{Z^1_{K^-}} \ ,
\label{eq:lamb-sig-k}
\end{eqnarray}
where $Z's$ indicate the different  one-particle partition functions
for  $K^+,K^-,\Sigma,\Lambda,\Sigma^*$.
In order to balance the
number of $K^+$, the main contribution in the $S=-1$ sector comes from
$\Lambda$ and $\Sigma$ hyperons and, in a smaller proportion, from
$K^-$ mesons and $\Sigma^*(1385)$ resonances.  On the other hand, the number of $K^-$ is
balanced by the presence of $K^+$ mesons. We finally observe that
the ratio is determined by the relative abundance of $\Lambda,\Sigma,\Sigma^*$
baryons with respect to that of $K^-$ mesons.

In order to introduce in-medium and temperature effects, the particles involved in the calculation should be dressed accordingly.
For $\Lambda$ and $\Sigma$, the partition
function reads
\begin{eqnarray}
Z_{\Lambda,\Sigma}=  g_{\Lambda,\Sigma}  V  \int
\frac{d^3p}{(2\pi)^3} e^{\frac{-\sqrt{m^2_{\Lambda,\Sigma}+p^2}-
U_{\Lambda,\Sigma}(\rho)+\mu_{B}}{T}} \ ,
\label{eq:lam-sig}
\end{eqnarray}
which is built using a mean-field dispersion relation for the
single-particle energies (see Refs.~\cite{Laura03,Balberg}),
while the resonance
$\Sigma^*(1385)$ is described by a Breit-Wigner shape.

With regards to the $K^-$ meson, two different prescriptions for the $K^-$ single-particle 
energy have been used. First, we
use the  mean-field approximation for the $K^-$
potential 
\begin{eqnarray}
Z_{K^-}&=& g_{K^-} V \int\frac{d^3p}{(2\pi)^3}
e^{\frac{-\sqrt{m_{K^-}^2 +p^2}-U_{K^-}(T,\rho,E_{K^-},p)}{T}} \ ,
\label{eq:onshell}
\end{eqnarray}
where $U_{K^-}(T,\rho,E_{K^-},p)$ is the $K^-$ single-particle
potential in the Brueckner-Hartree-Fock approach
\cite{Laura02}. 
The second approach incorporates the $K^-$ spectral
density (see Ref.~\cite{Laura03})
\begin{eqnarray}
Z_{K^-}&=& g_{K^-} V \int \frac{d^3p}{(2\pi)^3} \int ds
\ S_{K^-}(p,\sqrt{s}) \ e^{\frac{-\sqrt{s}}{T}} \ . 
\label{eq:offshell}
\end{eqnarray}

\section{Results}
%-----------------------------
\begin{figure}[htb]
\centerline{
\includegraphics[width=6.5cm,height=10cm,angle=-90]{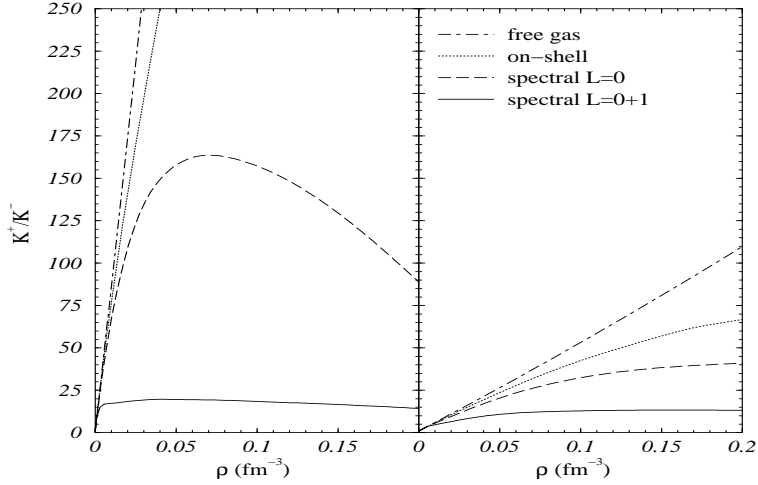}
}
\caption{$K^+/K^-$ ratio as a function of density for $T=50$ MeV
(left panel) and $T=80$ MeV (right panel) using different approaches
to the $K^-$ optical potential.}.
\label{fig:ratio-dens}
\end{figure}
%------------------------------------------
%-----------------------------------------------------------
\begin{figure}[htb]
\begin{minipage}
{150mm}
\centerline{
 \includegraphics[width=7cm]{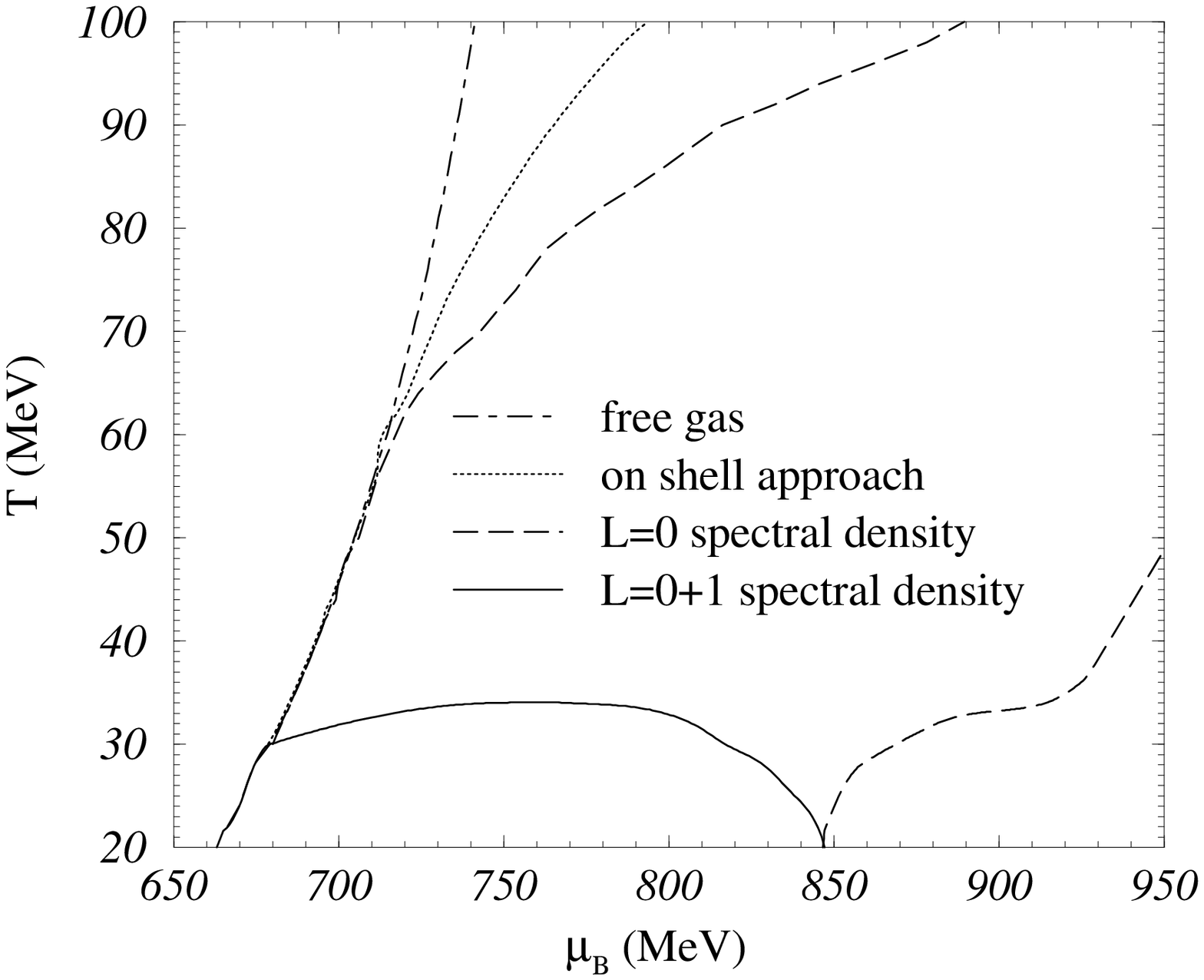}
\includegraphics[width=7cm]{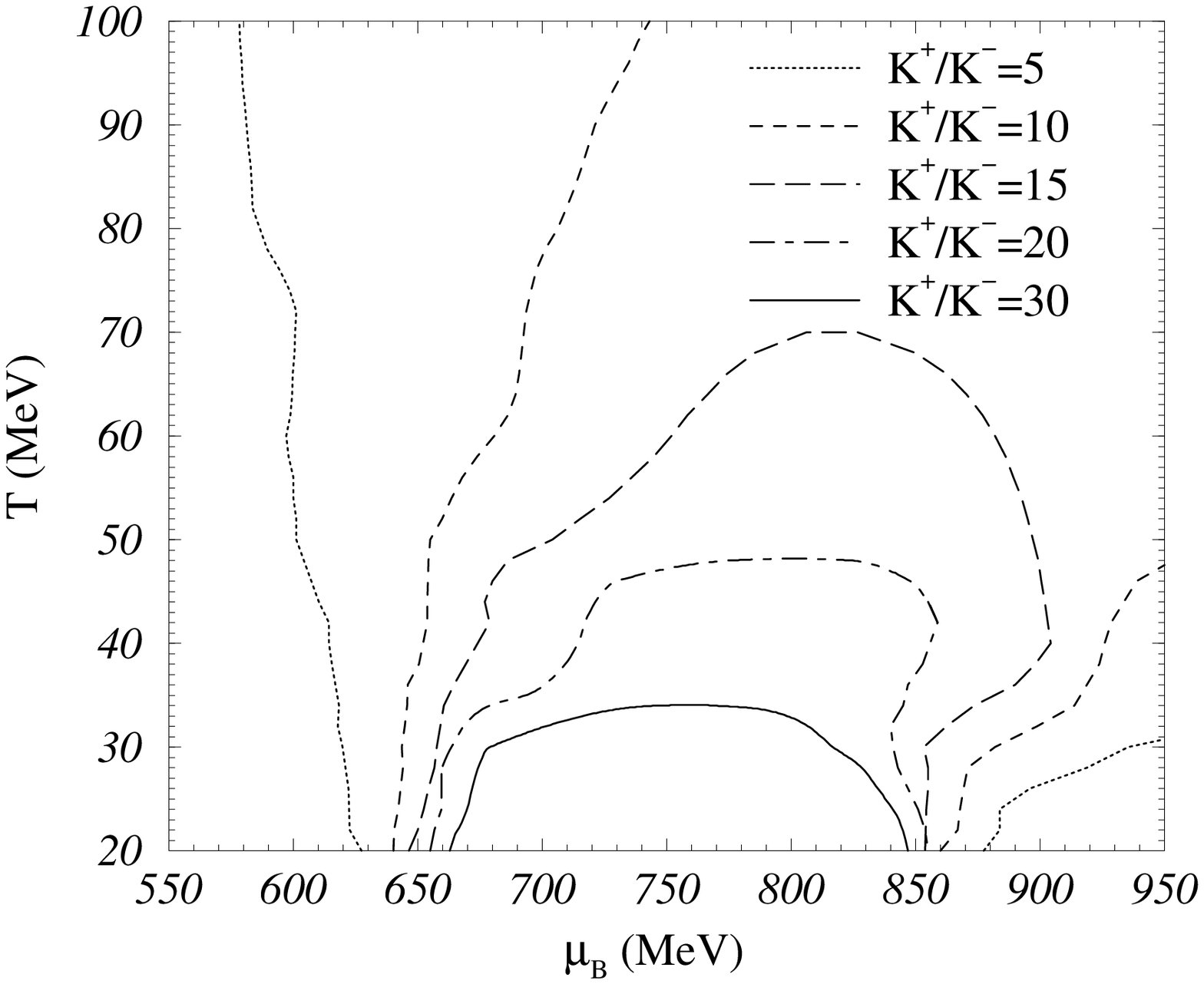}
}\caption{
Left: $T(\mu_B)$ for $K^+/K^-=30$ within different approaches. 
Right: Different $K^+/K^-$ ratios using the full $K^-$ spectral
density.}
\label{fig:t-m}
\end{minipage}
\end{figure} 

%------------------------------------------------------
In Fig.~\ref{fig:ratio-dens} the inverse ratio $K^+/K^-$ is shown
for two temperatures using different
approaches for the dressing of the $K^-$ meson: free gas (dot-dashed
lines), the on-shell approach (dotted lines) and using the $K^-$
spectral density including s-waves (long-dashed lines) or both s- and
p-waves (solid lines). The ratio grows with $e^{\mu_B/T}$
as we increase the density although it
 tends to bend down after the initial increase when the in-medium
$K^-$ properties are considered. Actually, when the full spectral density is used (solid lines), a flat region as a function of the density is observed. This is  due to the increased attraction produced by the  $YN^{-1}$ excitations  present in the low-energy region of the $K^-$ spectral density.  This result is in qualitative agreement with the
``broad-band equilibration'' advocated by Brown et al.\cite{Brown}.
However, this behaviour was found in Ref.~\cite{Brown} 
using a mean-field model through a
compensation of the increased attractive mean-field $K^-$
potential with the increase in the baryonic chemical potential as
density grows, which is not observed in our mean-field approach.

In the framework of statistical models, one obtains the relation between the temperature and the
chemical potential by fixing the $K^+/K^-$ ratio. In the
l.h.s. of Fig.~\ref{fig:t-m} we show, for the previous approaches,
 the values of temperature and
chemical potential compatible with a value of $K^+/K^-=30$, close to the experimental one for Ni+Ni collisions at 1.93 AGeV. Similar to the calculations of Refs.~\cite{Cleymans}, the dot-dashed lines gives the free gas case. While the on-shell
approach (dotted line) does not show the broad-band effect,  
 a band
of chemical potentials $\mu_B$ up to 850 MeV for $T
\approx 35$ MeV appears to be in accordance with the given ratio
 when both s- and p-wave contributions are
taken into account (solid line). However, in this case, the temperature
is too low to be compatible with the experimental one
and the corresponding freeze-out densities are too small. We can hardly speak of a ``broad band'' feature in the sense of that of Brown et al.  In the r.h.s.\ of
Fig.~\ref{fig:t-m} we display the temperature and chemical potential
for different values of the ratio when the full $K^-$ spectral density
is used. A ratio of the order of $15$ seems to be the solution for a more plausible experimental temperature of $T\approx 70$ MeV. 
We therefore conclude that the strength of the low-energy region of the $K^-$ spectral density gives an enhanced production of $K^-$ compared to the experimental results.

\section{Concluding remarks}

We have analyzed the $K^-$/$K^+$ ratio in the framework of  statistical models  considering the medium properties of antikaons. It is found that the determination of the temperature and baryonic chemical potential for a given ratio is very delicate depending on the approximation adopted for the $K^-$ self-energy. On the other hand,
the ``broad-band equilibration'' advocated by Brown, Rho and Song is
not accomplished in the on-shell approach.  Only when $K^-$ is described by the full spectral density we observe this broad-band. This is due to
the coupling of the $K^-$ meson to $YN^{-1}$ states.  However, the
$K^-/K^+$ ratio is in excess by a factor of 2 with respect to the
experimental value. Only further studies on non-equilibrium evolution of $K^-$ in the medium could give some indications about the reduced number of $K^-$ that are observed experimentally.

\vspace{-0.3cm}

\section*{Acknowledgments}
This work is partially supported by DGICYT project BFM2001-01868, by
the Generalitat de Catalunya project 2001SGR00064 and by NSF grant
PHY-03-11859. L. T. acknowledges support from the AvH Foundation.

\end{document}